\documentclass[12pt,reqno]{amsart}
\usepackage{graphicx,amsmath}
\usepackage[normalem]{ulem}
\usepackage{t1enc}              
\usepackage[utf8]{inputenc}    

\newcount\n  \newdimen\w

\def\Repeat#1#2{\n=#1\relax\loop\ifnum       
	\n>0\relax #2\advance\n by-1\repeat}

\long\def\OMIT#1{\relax }  

\def\re#1{(\ref{#1})}   
  
\def\eqn#1#2{ \begin{align} \label{#1}         #2 \end{align}}

\def\nl#1{          \\ \label{#1}        }  
\def\nnl#1{ \tag*{} \\ \label{#1}        }  
\def\nln#1{         \\ \label{#1} \tag*{}}  

\def\m#1{$            #1         $}  

\def\delim#1#2#3{\csname\ifcase#1 relax\or   
	big\or Big\or bigg\or Bigg\fi\endcsname   
	{\ifcase#2\or\Delim#3\or\deliM#3\fi}}      
\def\Delim#1{\ifcase#1\relax\or(\or[\or\{\or<\or\langle\or|\or\|\or---{ }\fi}
\def\deliM#1{\ifcase#1\relax\or)\or]\or\}\or>\or\rangle\or|\or\|\or{ }---\fi}

\let\f\frac                     
        
\def\largerfrac#1#2#3{      
	\whichtypesize\n=\currenttypesize\advance\n by #1 \mathchoice
	{\setbox0\hbox{$\displaystyle-$} \w=.5\ht0\advance\w by-.5\dp0\setbox0
		\hbox{\typesize\n $\displaystyle-$} \advance\w by -.5\ht0\advance\w
		by .5\dp0\raise\w \hbox{\typesize\n$\displaystyle{\frac{#2}{#3}}$}}
	{\setbox0\hbox{$-$} \w=.5\ht0 \advance\w by -.5\dp0 \setbox0\hbox
		{\typesize\n $-$} \advance\w by-.5\ht0\advance\w by
		.5\dp0\raise\w\hbox{\typesize\n$\frac{#2}{#3}$}}
	{\setbox0\hbox{$\scriptstyle-$} \w=.5\ht0 \advance\w by-.5\dp0\setbox0
		\hbox{\typesize\n $\scriptstyle-$} \advance\w by -.5\ht0 \advance\w
		by .5\dp0 \raise\w\hbox{\typesize\n$\scriptstyle{\frac{#2}{#3}}$}}
	{\setbox0\hbox{$\scriptscriptstyle-$} \w=.5\ht0
		\advance\w by -.5\dp0 \setbox0\hbox{\typesize\n
			$\scriptscriptstyle-$} \advance\w by -.5\ht0 \advance\w by .5\dp0
		\raise\w\hbox{\typesize\n$\scriptscriptstyle{\frac{#2}{#3}}$}}  }

\let\e\sle   

\def\bq{\mathbf x}
\def\bv{\mathbf v}
\def\ba{\mathbf a}

\def\re#1{(\ref{#1})}
\def\f{\frac}
\def\a{\mu} 	\def\i{i}  
\def\b{\nu} 	\def\j{j}
\def\c{\sigma} 	\def\k{k}
\def\d{\alpha} 	\def\l{l}
\def\e{\beta}  	\def\m{m}
\def\g{\gamma} 	\def\n{n}
\def\o#1{\overline{#1}}

\begin{document}
	
	\title{Generalized Galilean transformations of tensors and cotensors with application to general fluid motion}
	\author{V\'an P.$^{1,2,3}$, Ciancio, V,$^4$ and Restuccia, L.$^4$ }
	\address{$^1$HAS, Wigner Research Centre for Physics, Institute of Particla and Nuclear Physics, \\  
		and {$^2$Budapest University of Technology and Economics, Department of Energy Engineering},\\
		Montavid Thermodynamic Research Group\\
		$^4$ University of Messina, Department of Mathematical and Computer Sciences, Physical Sciences and Earth Sciences
		}
	
	\date{\today}

	
\begin{abstract}
	Galilean transformation properties of different physical quantities are investigated from the point of view of four dimensional Galilean relativistic (non-relativistic) space-time. The objectivity of balance equations of general heat conducting fluids and of the related physical quantities is treated as an application.

\end{abstract}
\maketitle
	
\section{Introduction}
	
	Inertial reference frames play a fundamental role in non-relativistic physics \cite{Lan14a,Pfi14a,PfiKin15b}. One expects that the physical content of spatiotemporal field theories are the same for different inertial observers, and also for some noninertial ones. This has several practical, structural consequences for theory construction and development \cite{Spe87c,Nol67a,NolSeg10a,Bre05a}. The distinguished role of inertial frames can be understood considering that they are true reflections of the mathematical structure of the flat, Galilean relativistic space-time \cite{Wey18b,Fri83b,Mat93b}. When a particular form of a physical quantity in a given Galilean inertial reference frame is expressed by the components of that quantity given in an other Galilean inertial reference frame, we have a Galilean transformation. In a four dimensional representation particular components of the physical quantity are called as timelike and spacelike parts, concepts that require further explanations. 
	
	 In non-relativistic physics time passes uniformly in any reference frame, it is absolute. 
	 The basic formula of Galilean transformation is related to both time and position:
	\eqn{Galtraf}{
		\hat t =t, \qquad \hat \bq = \bq - \bv t. 
	}
	
	\begin{figure}
		\centering
		\includegraphics[scale=0.5]{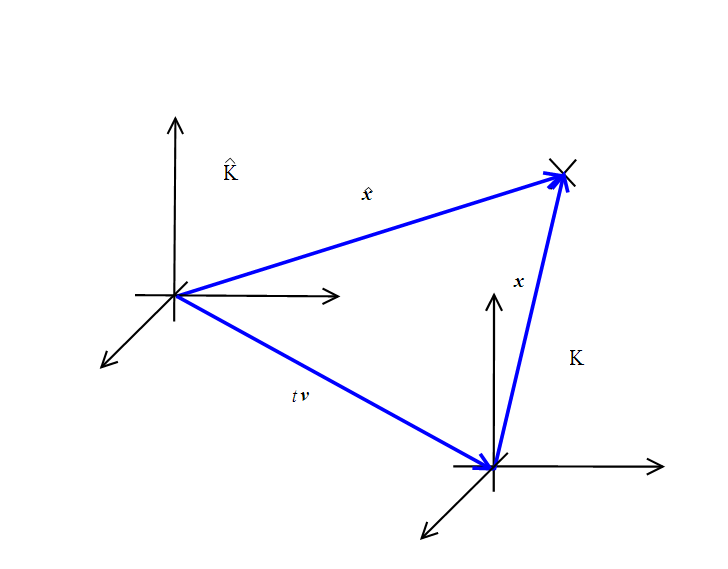}
		\caption{ \label{Gr_fig}
			Inertial reference frames $\hat K$ and $K$. }
	\end{figure}
	Here, $t$ is the time, $\bq$ is the position in an inertial reference frame $K$, $\hat\bq$ is the position in an other inertial reference frame $\hat K$ and $\bv$ is the relative velocity of $K$ and $\hat K$ (see Figure 1). The coordinates are not important parts of the reference frame, but there is a three dimensional Euclidean mathematical structure in both $K$ and $\hat K$. We can see from the transformation rule that time is absolute, it is the same in different inertial frames and position is relative, it is velocity dependent, the corresponding physical quantity changes when changing the reference frame. 
	
	In special relativity both space and time are relative, they are different for different inertial observers. There, time and space are not separate concepts, therefore, the natural physical quantities of four dimensional space-time are four-tensors. In non-relativistic physics one may think that due to absolute time, space is more separated from time, and the transformation rules influence only  three dimensional tensors. 
	 However, we will see, that several physical quantities are covectors \cite{Wes81b}, and for them not the spacelike, but the one dimensional timelike component is different in different inertial frames.
	
	Inertial reference frames seem to be distinguished for treating conceptual questions. However, it was recognized long time ago, that inertial reference frames are too special for continuum mechanics, basic equations are expected to be independent of the more general motion of rigid bodies, therefore objectivity requirements of physical quantities are formulated with transformation properties regarding changes of rotating reference frames \cite{Nol58a,TruNol65b}.  
	
	One may expect more: objective reality is expected to be independent of any external observer. Therefore physical theories -- the most objective models of reality -- are expected to be independent of reference frames. This is the ultimate reason, why we are investigating transformation rules. One may distinguish between three aspects of these investigations \cite{SveBer99a,Mus98a,MusRes02a,HerEta04p,MusRes08a,Mus12a,LiuSam14a}: 
	\begin{enumerate}
		\item The first is the reference frame dependence or independence of physical quantities. This is the question of {\em objectivity}.
		\item The second is the reference frame dependence or independence of laws of physics, in particular the objectivity of evolution equations and time derivatives.  This can be called the problem of {\em objective evolution}.
		\item The third is the  reference frame dependence or independence of material properties. This is the question of {\em material frame indifference}.
	\end{enumerate}	
	
	These aspects are not independent. One may expect that a frame independent representation of space-time automatically fulfils the above mentioned aspects \cite{Mur03a,NolSeg10a}. A realization of these ideas is given by the precise space-time models of Matolcsi \cite{Mat84b,Mat93b,Mat84b}. In this paper we mostly analyse the first question, the  objectivity of physical quantities and also pay attention to the second one and shortly mention the objectivity of evolution equations, in particular the basic balances of continua. In this work we do not develop a completely frame independent formalism as it was accomplished for Galilean relativistic fluids in \cite{Van15m1}. Our devices are inertial reference frames in Galilean relativistic space-time. In the first section we argue that physical quantities in non-relativistic physics are best considered as Galilean relativistic four-tensors, and their space- and timelike components cannot be separated. Here, our basic observation is that four-vectors and four-covectors transform differently. Then, we focus on Galilean transformations and derive the transformation rules of different four tensors, and provide some physical examples. The obtained transformation rules are similar to the ones obtained by Ruggeri assuming Galilean invariance of the systems of balances in Rational Extended Thermodynamics 
	\cite{Rug89a,MulRug98b,RugSug15b}. Finally, we discuss the results.

\section{Four-quantities in Galilean relativistic space-time: vectors and covectors}

In relativistic physics, both in special and general relativity, one introduces four-vectors and four-tensors, genuine four dimensional physical quantities, because there both time and space depend on the motion of the observer. The reference frame independent, absolute time of non-relativistic physics is exchanged to absolute  propagation speed of electromagnetic radiation.  The observer is that splits the objective spacetime to relative time and relative space. The relative velocity of the observer and the spacetime event plays  a particular role, as it is clear from the transformation rules of Lorentz between the times and spaces of two observers:
\eqn{Lortraf}{
 \hat t = \gamma (t + \bv\cdot \bq), \qquad 
 \hat \bq = \bq+ \gamma\left(-\bv t+ \f{\gamma}{\gamma +1} (\bv \cdot \bq) \bv \right)
}
Here  the instant $\hat t$ and position $\hat\bq$ of an event with respect to an inertial observer is related of the instant $t$ and position $\bq$ of the same event respect to a different inertial observer. The two observers move with relative velocity $\bv$ with respect to each oterh. $\gamma = \f{1}{\sqrt{1-v^2}}$ is the Lorentz factor, and \re{Lortraf} is  expressed in natural units, where the speed of the light $c=1$. One can see that the transformation rule is not simple, also its derivation requires a careful argumentation (e.g. the usual simple figure of coordinate systems with parallel axes is misleading: there is no parallelism concept ahead of transformation rules \cite{Mol72b,Mat93b}). This is one of the reasons why textbooks do not give the complete formula.  One can see that both space and time are transformed, therefore it is easy to accept that a covariant, objective treatment is based on spacetime vectors and tensors. Four-quantities are far more practical treating fundamental questions in relativistic situations. 

With the reference frame independent absolute time of the Galilei transformation \re{Galtraf} one may have the impression that four-quantities are unnecessary in non-relativistic physics. Their use may be a mathematical convenience, but they are not necessary for physical considerations. On the other hand one can recognize physical relation of four-dimensional spacetime quantities without any knowledge of special relativity, for that purpose Galilean transformations are sufficient. 


The most important evolution equations of classical continua are the balance equations of extensive physical quantities. The basic fields of a balance are the density, $a$, and the current density,  $\ba$, of the corresponding physical quantity.  The density is a scalar, the current density is a three dimensional spacial vector. It is easy to see that its Galilean transformation rule is $\hat \ba =\ba+ a \bv$, where the current density in a given reference frame is different that in an other one. The difference is the convective part of the total current density, $a \bv$, where $\bv$ is the relative velocity of the inertial frames. Densities and current densities are transforming like time and position, there is a complete Galilean transformation of a four vector, analogous to \re{Galtraf}:
\eqn{Exttraf}{
\hat a  = a, \qquad \hat \ba = \ba+ a \bv.	
}
Therefore $a$ and $\ba$ are components of a four vector in the Galilean relativistic spacetime. At this point it is worth to introduce a different, more convenient notation with indices. Writing Galilean transformation in a tensorial form 
\eqn{gtrs}{
\hat x^0	=  x^0, \qquad  \hat x^\i	=  x^\i + v^\i x^0,
}
with $\hat x^0 = \hat t$ and  $ x^0 = t$ and $\i =1,2,3$, we have also a matrix form
\eqn{xtr}{
	\begin{pmatrix}	\hat x^0\\ \hat x^\i \end{pmatrix} = 
	\begin{pmatrix} x^0\\ x^\i+x^0 v^\i \end{pmatrix} = 
	\begin{pmatrix} 1 & 0_j\\ v^\i &\delta^\j_{\;\;\j} \end{pmatrix} 
	\begin{pmatrix} x^0\\ x^\j \end{pmatrix},
}
or equivalently, with a four-index notation, $\hat x^\a = {\mathcal A}^\a_{\;\;\b}x^\b$, where the indexes are $\a = (0,\i)$ and  $\b = (0,\j)$ and \eqn{trformr}{
	{\mathcal A}^\a_{\;\;\b} = \begin{pmatrix}
		{\mathcal A}^0_{\;\;0} & 	{\mathcal A}^0_{\;\;\j} \\ 	
		{\mathcal A}^\i_{\;\;0}& 	{\mathcal A}^\i_{\;\;\j}
	\end{pmatrix} =	
	\begin{pmatrix} 1 & 0_j\\ v^\i &\delta^\j_{\;\;\j} \end{pmatrix} .
}
In the following Greek indices take the values $0,1,2,3$ and latin indices take the values $1,2,3$. From tensor calculus for a four-dimensional vector $A^\a$ we have the following transformation law 
\eqn{trl}{
	\hat A^\a = A^\b \f{\partial \hat x^\a}{\partial  x^\b},
	\qquad \text{or} \qquad 
	\hat A^\a = F^\a_{\;\;\b} A^\b,
	\qquad \text{being} \qquad 
	F^\a_{\;\;\b} = \f{\partial \hat x^\a}{\partial  x^\b}.
}
 Because of Galilean transformation is linear we have 
\eqn{AFform}{
{\mathcal A}^\a_{\;\;\b} = F^\a_{\;\;\b} =	
	\begin{pmatrix} 1 & 0_j\\ v^\i &\delta^\j_{\;\;\j} \end{pmatrix},
}
and therefore $F^\a_{\;\;\b}$ is the Jacobian of this transformation. Following the above formula one can see that \re{Exttraf} is the transformation rule for $\hat a^\a = (\hat a^0,\hat a^\i)$. Here the three-vector $\ba$ in \re{Exttraf} has to  be denoted by $a^i$.  We will use the following letters analogously, $i,j,k = {1,2,3}$. Therefore the Galilean transformation rule  above can be written equivalently as
\eqn{Exttrafi}{
	\hat a^0  = a, \qquad \hat a^i = a^i+ a v^i.	
}

We will denote the four-vectors with four indices. For example the time-position four-vector of an event is $x^\a$ and the density-current four-vector is $A^\a$, where we have introduced the abstract four-indexes $\a,\b,\c \in\{0,1,2,3\}$. Let us emphasize again, that both the three and the four indices do not refer to particular coordinates, they are abstract in the Penrose sense \cite{Pen04b,Wal84b}, as it is customary in relativistic theories.  

A four-vector has its timelike and spacelike components $A^\a= (a^0,a^i)$ in one reference frame and a different representation in a different inertial reference frame is denoted by $\hat A^\a= (\hat a^0,\hat a^i)$. The interested reader may find the exact mathematical representation of Galilean-relativistic spacetimes e.g. in \cite{Mat93b} and a developed index notation in \cite{Van15m1}.

A scalar valued linear mapping of a vector is a four-covector, an element of the dual of the original linear space. One can calculate the transformation rule of a four-covector from this defining property. Denoting the covector with a lower index, as $B_\a$ and its time- and spacelike components as $B_\a = (b_0,b_i)$  the duality mapping is written as $B_\a A^\a$. This product is a scalar, must be invariant, therefore 
\eqn{insp}{
	B_\a A^\a = \hat B_\a \hat A^\a.
}
Then from \re{trl}/1 follows
\eqn{dtrl}{
	\left(\hat B_\a - B_\b \f{\partial  x^\b}{\partial\hat  x^\a}. \right)\hat A^\a = 0,
	}
and so 
\eqn{ctrl}{
	\hat B_\a = B_\b \f{\partial x^\b}{\partial \hat  x^\a},
	\qquad \text{or} \qquad 
	\hat B_\a = G_\a^{\;\;\b} B_\b,
	\qquad \text{being} \qquad 
	G_\a^{\;\;\b} = \f{\partial x^\b}{\partial \hat  x^\a}= (F^{-1})_\a^{\;\;\b}.
}
That is, according to the transformation rule of the four-vector:
\eqn{trderiv}{
 	B_\a A^\a =& \begin{pmatrix}b_0,b_\i\end{pmatrix}
 	\begin{pmatrix}	a^0\\a_\i \end{pmatrix} = 
 	a^0b_0 + b_\i a^\i = 
 	\hat B_\a \hat A^\a = 
 	\hat a^0\hat b_0 + \hat b_\i\hat a^\i = \nln{he}
 	& a^0 \hat b_0+  \hat b_\i (a^\i + a^0 v^\i) = 
 	a^0(\hat b_0 - \hat b_\i v^\i) +\hat b_\i a^\i.
}

Comparing the too sides one obtains the transformation rule of a covector:
\eqn{covtr}{
  \hat b_0 = b_0- v^\i b_\i, \qquad \hat b_\i = b_\i.	
}
That is the spacelike part of a covector does not transform, but the timelike part transforms. 

An example of four-covector is the four-derivative operator. That can be demonstrated by the differentiation of a  scalar field, a scalar valued space-time function,  $f(t,x^i)$. If this function represents a physical quantity, its value must be the same before and after a Galilean transformation, that is $f(\hat t,\hat x^i)=f(t,x^i)$. Applying the derivative operators $\partial_t$ and  $\partial_\i$ to $f$, and taking into account the inverse  of the transformations \re{gtrs}, we obtain
\eqn{dtr}{
	\f{\partial f}{\partial \hat x^i} = 	
	\f{\partial f}{\partial x^i}, \qquad \text{and}\qquad
		\f{\partial f}{\partial \hat t} = 	
		\left( -v^i\f{\partial }{\partial  x^i}  +	\f{\partial }{\partial t}\right) f . 	
	}

Therefore the four-derivative is a covector operator, because it transforms, according to the covector transformation rule \re{covtr}, that is $\partial_\a = G_\a^{\;\;\b}\hat\partial_\b$, where $G_\a^{\;\;\b} =(F^{-1})_\a^{\;\;\b}$ is
\eqn{Gtform}{
	G_\a^{\;\;\b} = 
	\begin{pmatrix}	G_0^{\;\;0} & G_0^{\;\;\j} \\ G_\i^{\;\;0} & G_\i^{\;\;\j} \end{pmatrix} = 
	\begin{pmatrix}	1 & -v^\j \\ 0_\i & \delta_\i^{\;\;\j} \end{pmatrix}.	
}

 A vector  balance is a contracted mapping of the spacetime derivative on a four vector (or tensor). It would be a four-divergence in an Euclidean space but that is not necessary from a mathematical point of view. We will write it in the following form:
\eqn{4bal}{
  \partial_\a A^\a = \sigma.	
} 
Here the source density term, $\sigma$, is an invariant scalar, and the time- and spacelike components of the four-derivation are $\partial_\a = (\partial_t, \partial_\i)$. The four derivation is a covector. An application of these transformation rules for both components of a four divergence gives
\eqn{baltraf}{
	\hat\partial_\a \hat A^\a = & \begin{pmatrix}\hat\partial_t,\hat\partial_\i\end{pmatrix}
	\begin{pmatrix}
		\hat a^0\\\hat a_\i
	\end{pmatrix} = 
	\hat\partial_t a^0+ \partial_\i \hat a^\i = 
	(\partial_t-v^\i\partial_\i)a^0 + \partial_\i(a^\i+a^0 v^\i ) = \nnl{hu}
	& \partial_t a^0 + a^0\partial_\i v^\i +\partial_\i a^\i = \sigma.
	}
Let be the comoving reference frame where the quantities are denoted without hat and a local inertial frame is denoted with hat. Then, one can recognize that $\partial_t = \hat \partial_t + v^\i\partial_\i$ is the substantial time derivative and $\hat a^\i = a^\i +a^0 v^\i$ is the total current density of $a^0$, which is decomposed in conductive current density, $a^\i$, and the convective current density $+a^0 v^\i$, as usual. The later one is sometiems called the flux of $a^0$, (see e.g. \cite{Gya70b}). The relation between the substantial and partial derivatives is the Galilean transformation of a timelike covector component.

Therefore, in treating Galilean transformations one cannot restrict himself to scalar looking timelike and Galilean transforming three dimensional spacial components: transformation rules of vectors and covectors are entirely different. In special relativity the transformation rules of vectors and covectors look like are almost the same, both time- and spacelike components are velocity dependent. In Galilean  relativity the four-vector formalism is enforced by the different vector-covector transformations. A mathematical analysis of the physical observations reveals, that unlike the special relativity where the Lorenz-form is a pseoudeeuclidean structure on spacetime level, in Galilean relativity only a linear structure is required by physical considerations: time and space are not connected by any metric structure \cite{Fri83b,Mat84b,Mat93b}.

It is straightforward to summarize the transformation formulas in a four matrix notation. The Galilean transformation rule of a vector is

\eqn{vtrafom}{
	\hat A^\a  =
	\begin{pmatrix}	\hat a^0\\ \hat a^\i \end{pmatrix} = 
	\begin{pmatrix} a^0\\ a^\i+a v^\i \end{pmatrix} = 
	\begin{pmatrix} 1 & 0_j\\ v^\i &\delta^\j_{\;\;\j} \end{pmatrix} 
	\begin{pmatrix} a^0\\ a^\j \end{pmatrix}  =  F^\a_{\;\;\b} A^\b.
}
The transformation rule of a covector is (see \re{ctrl}):
\eqn{cvtrafom}{
	\hat B_\a  = 
		\begin{pmatrix} \hat b_0 & \hat b_\i \end{pmatrix} =
		\begin{pmatrix} b_0- b_\j v^\j & b_\i \end{pmatrix}  = 
 	\begin{pmatrix} b_0 & b_\j \end{pmatrix} 
	\begin{pmatrix} 1 & 0^i\\ -v^\j &\delta^\j_{\;\;\i} \end{pmatrix}   =  
	 B_\b G^\b_{\;\;\a}.
}
Let us observe, that the second order four-tensors of the Galilean transformation are mixed, they have both contravariant and covariant components.

\section{Galilean transformation of  second order tensors, cotensors and mixed tensors}

The Galilean transformation rules of four-vectors and four-covectors in a Galilean relativistic spacetime lead to the Galilean transformation rules of higher order tensors and cotensors. 

\subsection{Second order four-tensors}

The timelike component of a second order four-tensor $T^{\a\b}$ will be a four-vector, $t^\a$, $(\a,\b=0,1,2,3)$, as well as the timelike component of its transpose $\overline{t}^\a$. The time-timelike component of the four-tensor is a scalar, identical with the timelike component of $t^\a$ and $\overline{t}^\a$ and it is denoted by $t$. The space-spacelike component is a second order three-tensor, $t^{\i\j}$, $(\i,\j=1,2,3)$, with $t^\j$ the time-spacelike component of  $T^{\a\b}$ and $\hat t^\i$ the space-time component of  $T^{\a\b}$. Therefore  $T^{\a\b}$ is represented by the following matrix form:
\eqn{2ten}{
T^{\a\b} = \begin{pmatrix}
	t & t^{\j} \\ \o{t}^\i & t^{\i\j}.
\end{pmatrix}	
}
Here we have used the following notation for the vector components: $t^\a = (t,t^\j)$ and $\overline{t}^\a = (t,\overline{t}^\i)$. 

With our four-index notation the Galilean transformation of a second order tensor  is $\hat T^{\a\b} = F^\a_{\;\;\d}F^\b_{\;\;\e}T^{\d\e} = F^\a_{\;\;\d}T^{\d\e} F_\e^{\;\;\b}$. Here the transpose of the tensor $F^\b_{\;\;\d}$ is denoted by changing the left-right order of the indices: $(F^*)^\b_{\;\;\d} = F_\d^{\;\;\b}$. Therefore, using the time and spacelike components one obtains a formula that can be calculated according to the rules of matrix multiplication
\eqn{Gtraf_2ten}{
	\hat T^{\a\b} =& 
	\begin{pmatrix}	\hat t & \hat t^{\j} \\  \hat {\o{t}^\i} & \hat t^{\i\j}
		\end{pmatrix}	=
	\begin{pmatrix} 1 & 0_l\\ v^j &\delta^j_{\;\;l} \end{pmatrix} 
		\begin{pmatrix} 1 & 0_k\\ v^i &\delta^i_{\;\;k} \end{pmatrix} \begin{pmatrix} t & t^{\l} \\ \o{t}^\k & t^{\k\l}\end{pmatrix}  \nnl{jo}
 	 =&\begin{pmatrix} 1 & 0_k\\ v^i &\delta^i_{\;\;k} \end{pmatrix}
	    \begin{pmatrix} t & t^{\l} \\ \o{t}^\k & t^{\k\l}\end{pmatrix} 
		\begin{pmatrix} 1 & v^j\\ 0_l  &\delta_l^{\;\;j} \end{pmatrix}  \nnl{jo2}
	 =& \begin{pmatrix} t & t^{\j}+t v^\j \\ 
		 	\o{t}^\i+t v^\i & t^{\i\j} + v^it^j + \o{t}^iv^j + tv^iv^j\end{pmatrix}.
}

\subsection{Second order cotensors}

The timelike component of a second order four-cotensor $R_{\a\b}$ seems to be a four-covector, $r_\a$, as well as the timelike component of its transpose $\overline{r}_\a$, however it is not true. It is best seen by the different the Galilean transformation properties. The time-timelike component of the four-cotensor is identical with the timelike component of $r_\a$ and $\o{r}_\a$ and it is denoted by $r$. The space-spacelike component is a second order three-cotensor, $r_{\i\j}$. Therefore  $R_{\a\b}$ may be represented by the following matrix form:
\eqn{2cten}{
	R_{\a\b} = \begin{pmatrix}
		r & r_{\j} \\ \o{r}_\i & r_{\i\j}
	\end{pmatrix}	
}

With our four-index notation the Galilean transformation of a second order cotensor is analogous to the second order four-tensor and is the following $\hat R_{\a\b} = G_\a^{\;\;\d} G_\b^{\;\;\e}R_{\d\e} = G_\a^{\;\;\d}R_{\d\e} G^\e_{\;\;\b}$. Therefore, using the time and spacelike components one obtains a formula that can be calculated according to the rules of matrix multiplication
\eqn{Gtraf_2cten}{
	\hat R_{\a\b} =& 
		\begin{pmatrix}
		\hat r & \hat r_{\j} \\ \hat {\o{r}_\i} & \hat r_{\i\j} \end{pmatrix}	\nnl{ju}
	=&\begin{pmatrix} 1 & -v^\k\\ 0_\i &\delta_\i^{\;\;\k} \end{pmatrix}
		\begin{pmatrix}	r & r_{\l} \\ \o{r}_\k & r_{\k\l} \end{pmatrix}
		\begin{pmatrix} 1 & 0_j\\ -v^\l  &\delta^\l_{\;\;\j} \end{pmatrix}  \nnl{ju2}
	=& \begin{pmatrix} r- {\o{r}_\k} v^\k - r_\l v^\l + v^\k v^\l r_{\k\l} & r_{\j}-
		v^\k r_{\k\j} \\ \o{r}_\i - r_{\i\l}v^\l & r_{\i\j}\end{pmatrix}. 
}
One can see, that the time-timelike component of a four-cotensor is one dimensional, but it transforms with Galilean transformation, therefore it is not a scalar quantity if invariance is expected. It transforms differently than the timelike component of a covector. Remarkable that space-spacelike component of a cotensor does not transform by Galilean transformation.

\subsection{Second order mixed tensors}

The Galilean transformation rules of second order mixed four-tensors are different from the ones for four-tensors or four-cotensors. In this subsection we calculate only the case when the first index is contravariant and the second index is covariant. We will see that the timelike component of $Q^\a_{\;\;\b}$ is not a four-vector regarding its transformation rules, but $q^\a$, the timelike component of its transpose is a four-cotensor  in this sense. The time-timelike component and the second order mixed three-cotensor space-spacelike component are denoted by $q$ and  $q^\i_\j$, respectively. The time-spacelike and space-timelike components are $q^\i$ and $q_\j$. Therefore  $Q^\a_{\;\;\b}$ may be represented by the following matrix form:
\eqn{2mten}{
	Q^\a_{\;\;\b} = \begin{pmatrix}
		q & q_{\j} \\ q^\i & q^\i_{\;\j}
	\end{pmatrix}.	
}

The transformation formula is the combination of vector and covector transformation and can be written with the help of our four-index notation in the following form $\hat Q^\a_{\;\;\b} = F^\a_{\;\;\d}G^{\;\;\e}_{\b}Q^\d_{\;\;\e} = F^\a_{\;\;\d}Q^\d_{\;\;\e}G^{\e}_{\;\;\b}$. With the rules of matrix multiplication it is presented as
\eqn{Gtraf_2mten}{
	\hat Q^\a_{\;\;\b} = & 
	\begin{pmatrix}
			\hat q & \hat q_{\j} \\ \hat q^\i & \hat q^\i_{\;\j} \end{pmatrix}	\nnl{ja}
	=&\begin{pmatrix} 1 & 0_\k\\ v^\i &\delta^\i_{\;\;\k} \end{pmatrix}
	\begin{pmatrix}		q & q_{\l} \\ q^\k & q^\k_{\;\l} \end{pmatrix}
	\begin{pmatrix} 1 & 0_j\\ -v^\l  &\delta^\l_{\;\;\j} \end{pmatrix}  \nnl{ja2}
	=& \begin{pmatrix} q - q_\k v^\k  & q_{\j} \\ q^\i +q v^\i - q^\i_{\;\;\l}v^\l -q_\l v^\l q^\i & q^\i_{\;\j}+ v^\i q_\j\end{pmatrix} 
}

\section{Galilean transformation of third order tensors, 1-2 and 2-1 mixed tensors and cotensors}

The previous methodology with matrix multiplication becomes more clumsy for higher order tensors. Nevertheless, with the transformation matrices one can easily find the transformation formulas of the different terms one by one.

\subsection{Third order (3,0) tensor}

A third order contravariant tensor, $Z^{\a\b\c}$, can be represented as a hypermatrix of its time- and spacelike components as follows:
\eqn{30ten}{
	Z^{\a\b\c} =	\begin{pmatrix}
		Z^{0\b\c} &  Z^{\i\b\c}
	\end{pmatrix} =
	 \begin{pmatrix}
		\begin{pmatrix}
			Z^{000} & Z^{00\k} \\ Z^{0\j 0} & Z^{0\j\k}
		\end{pmatrix}	
			&
		\begin{pmatrix}
			Z^{\i 00} & Z^{\i 0\k} \\ Z^{\i\j 0} & Z^{\i\j\k}
		\end{pmatrix}	
	\end{pmatrix},	
}

where we may write also
\eqn{3tenrep}{
	Z^{\a\b\c} =	\begin{pmatrix}
		Z^{0\b\c} \\  Z^{\i\b\c}
	\end{pmatrix} =
		\begin{pmatrix}
			Z^{\a 0\c} \\  Z^{\a \j\c}
		\end{pmatrix} =
			\begin{pmatrix}
				Z^{\a\b 0} \\  Z^{\a\b\k}
			\end{pmatrix}.
}

We introduce a notation of the components of (1,3) representation of four-tensors: $F^0_{\;\;0} = 0,F^0_{\;\;\j} = 0^j,F^\i_{\;\;0} = v^i,F^\i_{\;\;\j} = \delta^i_{\;\;j}$. In matrix form it is represented as
\eqn{trformrn}{
	F^\a_{\;\;\b} = \begin{pmatrix}
				F^0_{\;\;0} & 	F^0_{\;\;\j} \\ 	F^\i_{\;\;0}& 	F^\i_{\;\;\j}
		\end{pmatrix} =	
					\begin{pmatrix} 1 & 0_j\\ v^\i &\delta^\j_{\;\;\j} \end{pmatrix},
}
(see \re{AFform}). The transformation formula is the following:
\eqn{30trafg}{
\hat Z^{\a\b\c} = F^\a_{\;\;\d}F^\b_{\;\;\e}F^\c_{\;\;\g}Z^{\d\e\g}.
}

Therefore the components of $\hat Z^{\a\b\c}$ can be given one by one, by using the component index notation, $\a = (0,\i)$,  $\b = (0,\j)$, $\c = (0,\k)$, $\d = (0,\l)$, $\e = (0,\m)$ and $\g = (0,\n)$.  Here $0$ denotes the timelike component and the latin indexes $\i,\j,\k,\l,\m,\n$ are for the spacelike ones. One can realize that the  previous matrix representations \re{30ten} or \re{3tenrep} require special attention for calculations. It is better to perform an automatic direct calculation:
\eqn{30traf}{
\hat Z^{000} &= F^0_{\;\;0}F^0_{\;\;0}F^0_{\;\;0}Z^{000}+     
	F^0_{\;\;0}F^0_{\;\;0}F^0_{\;\;\n}Z^{00\n}+
	F^0_{\;\;0}F^0_{\;\;\m}F^0_{\;\;0}Z^{0\m 0}+ 
	F^0_{\;\;0}F^0_{\;\;\m}	F^0_{\;\;\n}Z^{0\m\n} +	
 \nnl{s0}
&	F^0_{\;\;\l}F^0_{\;\;0}F^0_{\;\;0}Z^{\l 00}+
	F^0_{\;\;\l}F^0_{\;\;0}F^0_{\;\;\n}Z^{\l 0\n}+
	F^0_{\;\;\l}F^0_{\;\;\m}F^0_{\;\;0}Z^{\l\m 0}+ 
	F^0_{\;\;\l}F^0_{\;\;\m}F^0_{\;\;\n}Z^{\l\m\n}
 \nnl{s02}
=& Z^{000}, \nnl{s2}
\hat  Z^{\i 00} &=  v^\i  Z^{000} + Z^{\i 00} , \qquad
\hat  Z^{0\j 0} =  v^\j  Z^{000}+ Z^{0\j 0} , \qquad
\hat  Z^{00\k} =   v^\k  Z^{000} + Z^{00\k}, \nnl{s3}
\hat  Z^{\i\j 0} &=  Z^{\i\j 0} + v^\i  Z^{0\j 0} + v^\j  Z^{\i 00} 
		 + v^\i v^\j  Z^{\i\j 0}, \nnl{s30}
\hat  Z^{\i 0 \k} &=  Z^{\i 0 \k} + v^\i  Z^{00\k} + v^\k  Z^{\i 00} 
		 + v^\i v^\k  Z^{\i0 \k}, \nnl{s32}
\hat  Z^{0\j\k} &=  Z^{0\j\k} + v^\j  Z^{00\k} + v^\k  Z^{0\j 0} 
		 + v^\j v^\k  Z^{000}, \nnl{s4}
\hat  Z^{\i\j\k} &=  Z^{\i\j \k} + v^\i  Z^{0\j\k} + v^\j  Z^{\i 0\k} + 
		v^\k  Z^{\i\j 0} + v^\i v^\j  Z^{00\k} + v^\i v^\k  Z^{0\j 0} + v^\j v^\k  Z^{\i 00} + v^\i v^\j v^\k  Z^{000}. 
}

\subsection{(2,1) tensor}

In this case the transformation rule is the following:
\eqn{21trafg}{
	\hat Y_\a^{\;\;\b\c} = G^{\;\;\d}_\a F^\b_{\;\;\e} F^\c_{\;\;\g} Y_\d^{\;\;\e\g}
}

Then we need also the transformation formula of covectors.
\eqn{Gform}{
G_\a^{\;\;\b} = 
	\begin{pmatrix}	G_0^{\;\;0} & G_0^{\;\;\j} \\ G_\i^{\;\;0} & G_\i^{\;\;\j} \end{pmatrix} = 
	\begin{pmatrix}	1 & -v^\j \\ 0_\i & \delta_\i^{\;\;\j} \end{pmatrix}	
}

Therefore the components of $\hat Y_\a^{\;\;\b\c}$ are given one by one as follows
\eqn{21trafr}{
	\hat Y_0^{\;\;00} =& 
	G^0_{\;\;0}F^0_{\;\;0}F^0_{\;\;0}Y_0^{\;\;00}+ 				
	G^0_{\;\;0}F^0_{\;\;0}F^0_{\;\;\n}Y_0^{\;\;0\n}+
	G^0_{\;\;0}F^0_{\;\;\m}F^0_{\;\;0}Y_0^{\;\;\m 0}+	G^0_{\;\;0}F^0_{\;\;\m}F^0_{\;\;\n}Y_0^{\;\;\m\n}+ \nnl{y0}
&   G^\l_{\;\;0}F^0_{\;\;0}F^0_{\;\;0}Y_\l^{\;\;00}+
	G^\l_{\;\;0}F^0_{\;\;0}F^0_{\;\;\n}Y_\l^{\;\;0\n}+
	G^\l_{\;\;0}F^0_{\;\;\m}F^0_{\;\;0}Y_\l^{\;\;\m 0}+ 	 G^\l_{\;\;0}F^0_{\;\;\m}F^0_{\;\;\n}Y_\l^{\;\;\m\n} \nnl{y02}
	=& Y_0^{\;\;00}-v^\l Y_\l^{\;\;00}, \nnl{y2}
	\hat  Y_\i^{\;\;00} &=   Y_\i^{\;\;00}, \nnl{y21}
	\hat   Y_0^{\;\;\j 0} &= v^\j Y_0^{\;\;0 0} + Y_0^{\;\;\j 0} -  
		v^\l v^\j Y_\l^{\;\;00} -  v^\l Y_\l^{\;\;\j 0}, \nnl{y22}
	\hat   Y_0^{\;\;0\k} &=  v^\k Y_0^{\;\;0 0} +Y_0^{\;\;0\k} 	-  
		v^\l v^\k Y_\l^{\;\;00}-  v^\l Y_\l^{\;\;0\k} , \nnl{y3}
	\hat   Y_0^{\;\;\j\k}  &= Y_0^{\;\;\j\k}+ v^\k Y_0^{\;\;\j 0}+ v^\j Y_0^{\;\;0 \k} - 
		v^\l Y_\l^{\;\;\j\k} + v^2 Y_0^{\;\;\j\k} - v^\l v^\j Y_\l^{\;\;0\k}- v^\k v^\l Y_\l^{\;\;\j 0} - v^\k v^\j  v^l Y_\l^{\;\;00}, \nnl{y30}
	\hat   Y_\i^{\;\;\j 0} &=    v^\j Y_\i^{\;\;0 0} +Y_\i^{\;\;\j 0} , \qquad 
	\hat   Y_\i^{\;\;0\k} =   v^\k Y_\i^{\;\;0 0} + Y_\i^{\;\;0\k }, \nnl{y32}
	\hat   Y_\i^{\;\;\j\k} &=   v^\k v^\j Y_\i^{\;\;00} + v^\j Y_\i^{\;\;0\k}+ v^\k Y_\i^{\;\;\j 0}  + Y_\i^{\;\;\j\k}. 
}

\subsection{(1,2) tensor}

In this case the transformation formula is the following:
\eqn{12trafg}{
	\hat X^\a_{\;\;\b\c} = F_{\;\;\d}^\a G^\b_{\;\;\e} G^\c_{\;\;\g}X^\d_{\;\;\e\g}
}

Then we use the transformation matrices for vectors and covectors again.

Therefore the components of $\hat X^\a_{\;\;\b\c} $ are given one by one as follows
\eqn{12traf}{
	\hat X^0_{\;\;00} =& 
		F^0_{\;\;0}G^0_{\;\;0}G^0_{\;\;0}X^0_{\;\;00}+ 	
		F^0_{\;\;0}G^0_{\;\;0}G^\n_{\;\;0}X^0_{\;\;0\n}+		
		F^0_{\;\;0}G^\m_{\;\;0}G^0_{\;\;0}X^0_{\;\;\m 0}+ 
		F^0_{\;\;0}G^\m_{\;\;0}G^\n_{\;\;0}X^0_{\;\;\m\n}+ \nnl{x0}
	&F^0_{\;\;\l}G^0_{\;\;0}G^0_{\;\;0}X^\l_{\;\;00}+
	F^0_{\;\;\l}G^0_{\;\;0}G^\n_{\;\;0}X^\l_{\;\;0\n}+
	F^0_{\;\;\l}G^\m_{\;\;0}G^0_{\;\;0}X^\l_{\;\;\m 0}+ 	 F^0_{\;\;\l}G^\m_{\;\;0}G^\n_{\;\;0}X^\l_{\;\;\m\n} \nnl{x02}
	=& X^0_{\;\;00}-v^\m X^0_{\;\;\m 0} - v^n X^0_{\;\;0\n} + 
		v^m v^n X^0_{\;\;\m\n}, \nnl{x2}
	\hat  X^\i_{\;\;00} &=   v^\i  X^0_{\;\;00}-
		v^\i v^\n X^0_{\;\;0\n}-
		v^\i v^\m X^0_{\;\;\m 0}+
		v^\i v^\m v^\n X^0_{\;\;\m \n}+
		X^\i_{\;\;00}	- 
		v^\n  X^\i_{\;\;0\n}- 
		v^\m  X^\i_{\;\;\m 0}, \nnl{x21}
	\hat   X^0_{\;\;\j 0} &=  X^0_{\;\;\j 0} + v^\n  X^0_{\;\;\j \n},\qquad 
		\hat   X^0_{\;\;0 \k} =  X^0_{\;\;0\k} + v^\m  X^0_{\;\;\m\k},\nnl{x3}
	\hat   X^0_{\;\;\j\k}  &=  X^0_{\;\;\j\k}, \nnl{x30}
	\hat   X^\i_{\;\;\j 0} &=  v^\i X^0_{\;\;\j 0} - v^\i v^\n X^0_{\;\;\j \n}
		+ X^\i_{\;\;\j 0} - 		v^\n X^\i_{\;\;\j \n}   , \qquad  \nnl{x31}
	\hat   X^\i_{\;\;0\k} &=  v^\i X^0_{\;\;0\k}- v^\i v^\m X^0_{\;\;\m \k} + X^\i_{\;\;0\k}  - 
		v^\m X^\i_{\;\;\m \k} , \nnl{x32}
	\hat  X^\i_{\;\;\j\k} &=  v^\i X^0_{\;\;\j\k} + X^\i_{\;\;\j\k}. 
}

\subsection{Third order cotensor}
Finally we can give the following transformation rule of a third order cotensor:
\eqn{03trafg}{
	\hat W_{\a\b\c} = G_{\;\;\a}^\d G^\e_{\;\;\b} G^\g_{\;\;\c}W_{\d\e\g}
}

Then we use the proper combination of transformation formulas for vectors and covectors again. Therefore the components of $\hat W_{\a\b\c} $ are given one by one as follows
\eqn{03traf}{
	\hat W_{000} =& 
	G^0_{\;\;0}G^0_{\;\;0}G^0_{\;\;0}W_{000}+ 	
	G^0_{\;\;0}G^0_{\;\;0}G^\n_{\;\;0}W_{00\n}+ 	
	G^0_{\;\;0}G^\m_{\;\;0}G^0_{\;\;0}W_{0\m 0}+ 
	G^0_{\;\;0}G^\m_{\;\;0}G^\n_{\;\;0}W_{0\m \n}+
	\nnl{w0}	&
	G^\l_{\;\;0}G^0_{\;\;0}G^0_{\;\;0}W_{\l 00}+
	G^\l_{\;\;0}G^0_{\;\;0}G^\n_{\;\;0}W_{\l 0\n}+
	G^\l_{\;\;0}G^\m_{\;\;0}G^0_{\;\;0}W_{\l \m0}+ 	
	 G^\l_{\;\;0}G^\m_{\;\;0}G^\n_{\;\;0}W_{\l\m\n}, \nnl{w02}
	=& W_{000}-v^\l W_{\l 00}-v^\m W_{0\m 0}-v^\n W_{00\n}  +
		v^\l v^\m W_{\l\m 0}+  \nnl{w03}
		&\qquad v^\l v^\n W_{\l 0\n}+ v^\m v^\n W_{0\m\n} -
		v^\l v^\m v^\n W_{\l\m \n}, \nnl{w2}
	\hat  W_{\i 00} &=   W_{\i 00} - v^\m  W_{\i \m 0} - v^\n  W_{\i 0\n} +
		 v^\m v^\n  W_{\i \m \n},	 		\nnl{w21}
	\hat  W_{0 \j 0} &=  W_{0 \j 0} - v^\l  W_{\l \j 0} - v^\n  W_{0\j\n} +
		 	v^\l v^\n  W_{\l \j \n}, 	 	\nnl{w22}
	\hat  W_{00\k} &=  W_{00\k} - v^\l  W_{0 \l \k} - v^\m  W_{0\m\k} +
		 		v^\l v^\m  W_{\l \m \k}, 	\nnl{w23}
	\hat   W_{\i \j 0} &=  W_{\i \j 0} - v^\n  W_{\i \j \n},\qquad 
		\hat   W_{\i 0 \k} =  W_{\i 0\k} - v^\m  W_{\i \m \k},\qquad 
		\hat   W_{0\j \k} =  W_{0\j \k} - v^\l  W_{\l\j \k},\nnl{w3}
	\hat   W_{\i \j \k} &=  W_{\i \j \k}.
}

\section{Balances and identification of physical quantities}


Galilean transformation properties of higher order tensors are not treated frequently in the literature of continuum physics. According to the common knowledge  for non-kinematic physical quantities the transformation properties  are to be postulated \cite{LiuSam14a}. An exemption comes from Rational Extended Thermodynamics, where transformation properties of the fields are calculated with the requirement of Galilean invariance of the system of basic balances \cite{Rug89a,MulRug98b}. 

Here, our simple and basic assumption is the that the four divergence of a third order tensor field $Z^{\a\b\c}$, gives the system of basic balances. The physical justification is given with the help of the calculated transformation rules. 

Because the four-divergence of a third order tensor field  is a second order tensor, we also introduce the corresponding source term, as follows:
\eqn{3bal}{
\hat\partial_\a \hat Z^{\a\b\c} = \hat T^{\b\c}.
}

Here and in the following we assume that $Z^{\a\b\c}$ is symmetric in the last two indexes: $Z^{\a\b\c} = Z^{\a\c\b}$. In a time/spacelike decomposition one obtains easily the following system of balances:
\eqn{3bals}{
\hat\partial_t \hat Z^{000} &+ \partial_\i \hat Z^{\i 00} = \hat T^{00}\nnl{no1}	
\hat\partial_t \hat Z^{0\j 0}&+ \partial_\i \hat Z^{\i \j 0} = \hat T^{\j 0}\nnl{no2}	
\hat\partial_t \hat Z^{0\j \k}&+ \partial_\i \hat Z^{\i \j\k} = \hat T^{\j\k}		
}

With the help of the previously calculated  transformation rules follows that
\eqn{mbalinv}{
	\partial_t& Z^{000} +Z^{000}\partial_\i v^i+ \partial_\i Z^{\i 00} =  T^{00}\nl{lbalinv}	
	\partial_t &(Z^{0\j 0}+ v^\j Z^{000})+ 
		(Z^{0\j 0}+ v^\j Z^{000})\partial_\i v^\i +
		\partial_\i( Z^{\i \j 0}+ v^\j Z^{\i 00}) =  T^{0\j}+v^j T^{00}  \nl{nr2}	
	\partial_t& (Z^{0\j \k}+ v^\j Z^{00\k}+ v^\k Z^{0\j 0}+ v^\j v^k Z^{0\j\k})+ 
	(Z^{0\j \k}+ v^\j Z^{00\k}+ v^\k Z^{0\j 0}+ v^\j v^k Z^{0\j\k})\partial_\i v^\i +\nnl{bf} 
	 &\partial_\i( Z^{\i\j \k} +  v^\j  Z^{\i 0\k} + 
	 v^\k  Z^{\i\j 0}  + v^\j v^\k  Z^{\i 00} + v^\i v^\j v^\k  Z^{000}) =  \nnl{ebalinv} &T^{\j\k}+v^\j T^{0\k}+v^\k T^{\j 0}+v^\j v^\k T^{00}.	
}

There one may introduce the following notation: $\rho = Z^{000}$, $j^\i = Z^{\i 00}$,  $\sigma =T^{00}$ and we denote the substantial time derivative $\partial_t$ by an overdot. Then \re{mbalinv} is apparently the  balance of mass, the well know equation of continuity in a local form, where $\rho$ is the mass density and $v^i$ is the local velocity field:
\eqn{balm}{
\dot \rho + \rho \partial_\i v^\i+  \partial_\i j^\i= \sigma.
}
Here $j^i$ is a conductive mass flux, a (self)diffusion term.  Usually one assumes, that  $j^i$ and also the source term $\sigma$ is zero. In \cite{OttEta09a} and \cite{VanEta16a} conditions are given when it may happen. Booster conservation, the existence of local center of mass motion exclude nonzero (self)diffusion. 

The new components of $Z^{\a\b\c}$ in the second equation, \re{lbalinv}, will be denoted as $p^\j = Z^{0\j 0}$, $P^{\i\j} = Z^{\i\j 0}$ and  $f^\j = T^{0\j}$. Then it can be written in the following form
\eqn{bali}{
 (p^\j+ \rho v^\j\dot)+ (p^\j+ \rho v^\j)\partial_\i v^\i +
	\partial_\i(P^{\i\j}+ v^\i p^\j+ v^\j j^\i+ v^\i v^\j \rho) =  f^\j +v^\j \sigma
}
Here we can see, that if $p^\j$ and $j^\j$ are zero, then  one obtains a simplified form:
\eqn{balir}{
	 (\rho v^\j\dot)+ \rho v^\j\partial_\i v^\i + \partial_\i(P^{\i\j} + v^\i v^\j \rho) =  f^\j 
}

We can recognise the balance of momentum, with  $P^{\i\j}$ as pressure tensor and $f^\j$ as force density.  $p^\j$ is the density of the momentum, the part that does not flow with the fluid. 

Finally, let us take the trace of the second order tensor equation \re{ebalinv} and introduce the following notations: $Z^{0\j \j} = 2e$, $Z^{\i\j \j} = 2q^\i$ and also $T^{\j\j} = 2\Sigma $. Then
\eqn{bale}{
 (e+ &v^\j p^\j+ \f{\rho}{2} v^\j v^\j \dot)+ 
 (e+ v^\j p^\j+ \f{\rho}{2} v^\j v^\j )\partial_\i v^\i +
	\partial_\i( q^\i +  v^\k  P^{\i\k}  + \f{1}{2}v^\j v^\j  j^{\i}) =  \nnl{bf11} &\Sigma + v^\j f^{\j}+v^\j v\j \f{\sigma}{2},	
}
where $e$ is the internal energy density and $q^i$ is the heat flux vector.
Beyond the previous assumptions we consider the special case when the basic source term $\Sigma $ is zero and finally obtain, that
\eqn{baler}{
	(e+  \f{\rho}{2} v^\j v^\j \dot)+ 
	(e+ \f{\rho}{2} v^\j v^\j )\partial_\i v^\i +
	\partial_\i( q^\i +  v^\k  P^{\i\k}) = v^\j f^{\j}.	
}
Here, we recognize the balance of total energy.

With the above notation, and applying the symmetry of $Z$, the transformation rules \re{s4}  reduce to the following form:
\eqn{30trafph}{
	\hat \rho 	 	&= \rho, \nnl{sph2}
	\hat  j^{\i} 	&=  j^{\i} + v^\i  \rho, \qquad
		\hat  p^{\j} =  p^{\j} + v^\j \rho, \nnl{ss3}
	\hat  P^{\i\j} 	&=  P^{\i\j} + v^\i  p^{\j} + v^\j  j^{\i} + v^\i v^\j  \rho,\nnl{ss30}
	\hat e 			&=  e + v^\k  p^{\k} +  v^2  \f{\rho}{2},\nnl{ss4}
	\hat  q^{\i} 	&=  q^{\i} + v^\i  \hat e + v^\k  P^{\i\k} + j^\i v^2 \f{\rho}{2}.
}

It is worth to observe that without the mentioned simplifying assumptions $p^\i= 0$, $j^\i= 0$, the transformed form of the current densities $P^{\i\j}$ and $ q^{\i} $ are the usual total form compared to the convective and convective ones. On the other hand, we can see that the conditions $p^\i= 0$ and $j^\i= 0$ can be considered as defining special reference frames, moving together with the mass and the momentum.It is hard to recognize the transformation rules without these annulled terms. It is remarkable that the kinetic energy represents a part of the transformation rule of the energy.

\section{Summary and discussion}

In this paper we have calculated the Galilean transformation rules of the basic second and third order four-tensors in Galilean relativistic spacetime.  Then these transformation rules -- with a local interpretation, of non-constant transformation velocity -- were fitted into the balances in order to  realize the physical meaning and relevance of the component physical quantities more easily.

According to the present work the basic balances of continua have the same meaning in inertial reference frames. The mass, momentum and energy balance in the presented form are not invariant but transform according to Galilean transformations. The particular transformation rules are derived and we have obtained the same formulas as one can obtain in Rational Extended Thermodynamics. There the derivation is based on the Galilean invariance of the whole system of balances. Here the transformation rules reveal a four-tensor structure of the basic physical quantities of classical fluid mechanics. 

It is also remarkable that the treated laws of physics are not  invariant when changing inertial reference frames, however, they transform according to proper transformation rules. The concept of objectivity and objective evolution are the same, a simple four-dimensional representation proves this identity. One may realize that our demonstration here is applicable to general fluid motions, with generalized Galilean transformations, where $v^\i$ is the relative velocity of the fluid and an inertial observer, therefore it is not necessarily constant. An other instructive example is given in \cite{MatVan06a} for Euclidean transformations and with inertial forces.

Some of these components are considered as zero in the classical treatments. The physical existence and nonexistence conductive  mass flux $j^\i$ was recently discussed in the literature \cite{Bre05a,BedEta06a,OttEta09a,VanEta16a}. The (self)momentum density $p^i$ does not appear in case of fluids. Their common zero value means that the mass and the momentum of the fluid flow together, an assumption that seems to be evidently true in case of  massive pointlike, classical particles. 

This treatment is a particular case of a completely reference and flow-frame independent approach, more general but less transparent from the presented Galilean transformations based treatment \cite{Van15m1}. 

Our presented approach is admittedly  Pythagorean, where the mathematical form precedes the physical representations.

\section{Acknowledgement}   

This work was supported by the grants Otka K104260 and K116197 and by University of Messina Grant.

\bibliographystyle{unsrt}

\end{document}